\documentclass[prl,twocolumn]{revtex4-1}

\usepackage{graphicx,epsfig,epstopdf,color}
\usepackage{amsmath,amssymb,bm}

\newcommand{\lap}{\nabla^2}

\newcommand{\Br}{{\bf r}}

\begin{document}

\title{Imaging from the Inside Out: \\
 Inverse Scattering with Photoactivated Internal Sources} 

\author{Anna C. Gilbert}
\affiliation{Department of Mathematics, University of Michigan, Ann Arbor, MI 48109, USA}
\email{annacg@umich.edu}

\author{Howard W. Levinson}
\affiliation{Department of Mathematics, University of Michigan, Ann Arbor, MI 48109, USA{}}
\email{levh@umich.edu}

\author{John C. Schotland}
\affiliation{Department of Mathematics and Department of Physics, University of Michigan, Ann Arbor, MI 48109, USA}
\email{schotland@umich.edu}

\begin{abstract}
We propose a method to reconstruct the optical properties of a scattering medium with subwavelength resolution. The method is based on the solution to the inverse scattering problem with photoactivated internal sources. Numerical simulations of three-dimensional structures demonstrate that a resolution of approximately $\lambda/25 $ is achievable.
\end{abstract} 

\maketitle

Scattering experiments remain among the most powerful tools to probe the optical properties of matter. It is well known that the data obtained from such experiments contains information about the structure of material media~\cite{Born-Wolf}. Inverse scattering theory is concerned with the problem of recovering three-dimensional structure (in the form of the scattering potential) from scattering data. The usual statement of the inverse scattering problem (ISP) is to reconstruct the scattering potential from measurements of the optical field taken \emph{external} to the scatterer. The ISP can then be formulated in a variety of settings, depending upon the nature of the incident field and the method of detection. Regardless of such considerations, the fundamental questions to be investigated are the uniqueness, stability, and reconstruction of the solution to the inverse problem. We note that ISPs are typically ill-posed, which means that they must be suitably regularized to achieve stable inversion.  

Subwavelength resolution (also known as superresolution) can be achieved by making use of near-field measurements~\cite{novotny}. This principle has been exploited in near-field microscopy and in several related tomographic imaging modalities~\cite{carney_2001,carney_2004,govyadinov_2009}. The corresponding inverse problems are severely ill-posed due to the exponential decay of evanescent waves. As a result, only relatively limited improvements in resolution are achievable in practice~\cite{govyadinov}.
 
The availability of \emph{internal}, rather than boundary measurements of the optical field would fundamentally alter the mathematical approach to the ISP. In particular, the scattering potential could then be obtained by a stable local computation~\cite{remark}. Unfortunately, the necessary measurements of the optical field cannot be acquired in practice. However, the principle of reciprocity implies that equivalent information can be obtained by replacing internal detectors with internal sources. Suitable sources are available in the form of photoactivatable fluorescent molecules. Such molecules are the basis of photoactivated localization microscopy (PALM), a form of superresolution fluorescence microscopy with spatial resolution on the order of ten nanometers~\cite{Betzig1642,Patterson,Roadmap}. In PALM, the image of a sample is built up sequentially from images that are recorded when only a small number of well-separated fluorophores are activated. The position of a fluorophore is obtained by estimating the centroid of the corresponding point spread function. In this manner, a single fluorescent molecular can be localized with subwavelength accuracy, despite the fact that the measurement is carried out in the far-field. As a consequence, the resolution of the acquired image is limited by the precision of localization of the individual fluorescent source molecules, rather than by diffraction. Despite this remarkable property, the PALM image is not related to the optical properties of the medium. Rather, it is a map of the fluorescent molecules with which the sample is labeled.

In this Letter we investigate the ISP with internal sources. We show that it is possible to stably reconstruct the scattering potential from PALM measurements.  Since the sources are in the near-field of the scatterer, our approach enables the recovery of the scattering potential with subwavelength resolution. 
The added value of the proposed method to PALM is that the reconstructed images are tomographic and are quantitatively related to the optical properties of the medium. The ready availability of photoactivated fluorophores, together with the well-posedness (stability) of the ISP with internal data is expected to allow the practical realization of the method.

We begin by considering an experiment in which light from a monochromatic point source propagates in a scattering medium. The source is taken to be located in the interior of the medium and to consist of a single photoactivated fluorescent molecule. The position of the source is assumed to be known, as determined from a PALM image. A series of such experiments is conducted, in which many fluorophores are activated one at a time. The resulting data set is then used to reconstruct the scattering potential of the medium. For simplicity, we ignore the effects of polarization and employ a scalar theory of the optical field. The field $u$ obeys the reduced wave equation
\begin{equation}
\label{wave_eqn}
\lap u + k^2\varepsilon(\Br) u = \delta(\Br-\Br_0) \ ,
\end{equation}
where $\varepsilon$ is the dielectric permittivity of the medium, $\Br_0$ is the position of the source and $k$ is the wavenumber. The field may be decomposed as the sum $u=u_i+u_s$, where $u_i$ is the incident field and $u_s$ is the scattered field. The incident field satisfies (\ref{wave_eqn}) in the absence of the scatterer. The scattered field obeys the integral equation
\begin{equation}
\label{LS}
u_s(\Br) =  \int_\Omega d^3r' G(\Br,\Br') V(\Br') u(\Br') \ ,
\end{equation}
where the outgoing Green's function $G$ is given by
\begin{equation}
\label{eq:G0}
G(\Br,\Br')=\frac{e^{ik|\Br-\Br'|}}{4\pi|\Br-\Br'|} \ ,
\end{equation}
the scattering potential $V$ is defined by $V(\Br)=k^2(\varepsilon(\Br)-1)$, and $\Omega$ is the volume of the scatterer.
Suppose that the medium is a microscopic weakly-scattering dielectric. We may then calculate the scattered field by making use of the first Born approximation (FBA) to linearize (\ref{LS})~\cite{Born-Wolf}. Noting that the incident field is related to the Green's function by $u_i(\Br)=G(\Br,\Br_0)$, we find that in the far-zone of the scatterer, the scattered field has the asymptotic form
\begin{equation}
u_s \sim \frac{1}{4\pi}\frac{e^{ik r_2}}{r_2} A(\Br_1,\Br_2) \ ,
\end{equation} 
where
\begin{equation}
\label{def_A}
A(\Br_1,\Br_2) = \int_\Omega d^3r e^{-ik\hat\Br_2\cdot\Br}G(\Br,\Br_1)V(\Br) \ .
\end{equation}
Here the dependence of the scattering amplitude $A$ on the source position $\Br_1$
and detector position $\Br_2$ has been made explicit. 

The inverse problem is to reconstruct the scattering potential $V$ from measurements of the scattering amplitude $A$. It is important to note that $A$ is generally complex-valued and that the phase of the scattered field can be obtained from interferometric PALM experiments~\cite{Shtengel03032009,Shtengel2014273}. To proceed, we must solve the integral equation (\ref{def_A}) that relates $V$ to $A$. One approach to this problem is to discretize the volume $\Omega$, thereby converting (\ref{def_A}) into a system of linear algebraic equations of the form $Kx=y$. Such equations must be regularized, which leads to the linear system
\begin{equation}
(K^*K+\alpha^2I)x=K^*y \ .
\end{equation} 
Here $K$, $x$ and $y$ correspond to the discretized form of the kernel $e^{-ik\hat\Br_2\cdot\Br}G(\Br,\Br_1)$, $V$ and $A$, respectively.  We solve the above system by conjugate gradient descent with the regularization parameter $\alpha$ chosen by the L-curve method~\cite{hansen}. 

\begin{figure}[t]
\centering
\includegraphics[height=4.95cm]{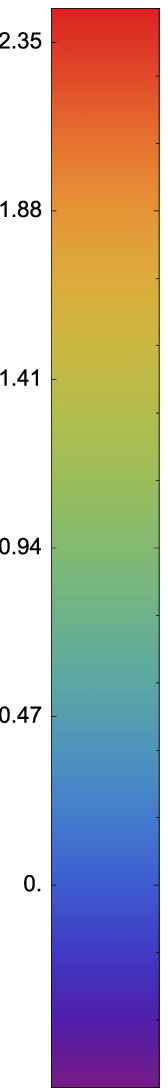}\includegraphics[width=.9\linewidth]{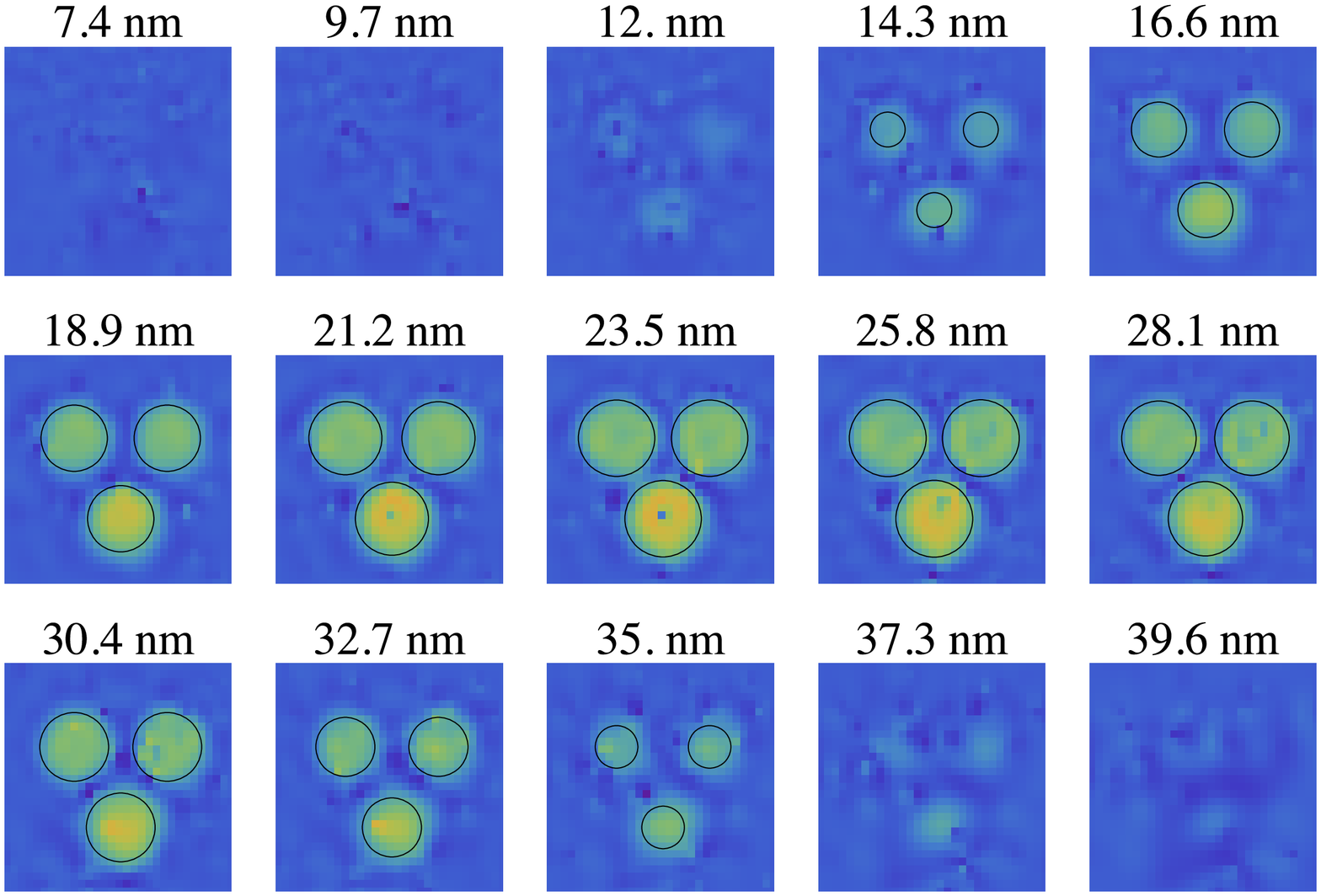}
\caption{(Color online) Reconstruction of three spherical scatterers in slices parallel to the image plane at the indicated depths. The black circular lines represent the true outline of the scatterers.  The field of view in each image is 70 nm $\times$ 70 nm.}
\label{fig:3Dslices}
\end{figure}

\begin{figure}[b]
\centering
\includegraphics[width=0.70\linewidth]{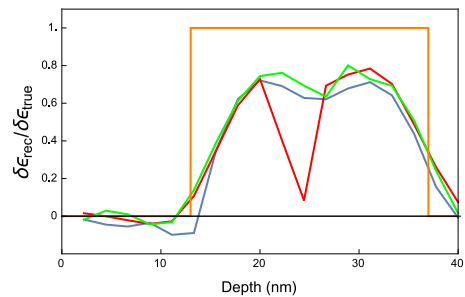}
\caption{(Color online) One-dimensional profiles through the centers of each spherical scatterer.  The quantity displayed is $\delta\varepsilon_{\rm rec}/\delta\varepsilon_{\rm true}$. The orange line represents the model, the blue and green lines correspond to the scattererers with $\delta\varepsilon=1.26$, and the red line represents the scatterer with $\delta\varepsilon=1.89$.}
\label{fig:3Dcs}
\end{figure}

To illustrate the reconstruction method, we have numerically simulated the reconstruction of a model system and a spiny neuron. The forward data was generated by solving (\ref{wave_eqn}) by the coupled-dipole method~\cite{43110825,TMatrix2}.  Gaussian white noise was added to the data as indicated below. The wavelength of light is $\lambda=2\pi/k = 500$nm. We have assumed that all fluorophores emit at the same wavelength, an assumption that is easily relaxed. The model system consisted of a volume of dimensions 70nm $\times$ 70nm $\times$ 40nm.  Three spherical scatterers of radius 12nm were placed so their lowest point was 3nm from the bottom of the sample. The spheres were placed so that the minimum distance between any two spheres was 5nm. The contrast level of two of the spheres was set to $\delta\varepsilon:=V/k^2=1.257$, while the remaining sphere had $\delta\varepsilon=1.885$.  The scattered field was registered on a 13 $\times$ 13 evenly spaced grid of detectors that spanned a 120 degree angular field of view in both lateral directions. In addition, 150 sources were randomly placed throughout the sample, corresponding to 25,350 measurements to which 1\% noise is added. The forward problem was solved on a grid consisting of 24,500 voxels with volume 8nm$^3$. The inverse problem was solved on a grid of 17,298 voxels with volume 12.17 nm$^3$, thereby avoiding so-called inverse crime. The reconstructions are displayed in Fig.~\ref{fig:3Dslices}, in which two-dimensional tomographic slices are shown. In Fig.~\ref{fig:3Dcs} we show the one-dimensional profiles of the reconstructions drawn through the centers of each scatterer. It can be seen that the spheres are clearly reconstructed with resolution of approximately $\lambda/25$, which corresponds to the FWHM of the curves shown in Fig.~\ref{fig:3Dcs}. We note that the support of the scatterers is recovered relatively accurately, but the reconstructed contrast is underestimated.

\begin{figure}[t]
\centering
\includegraphics[height=4.5cm]{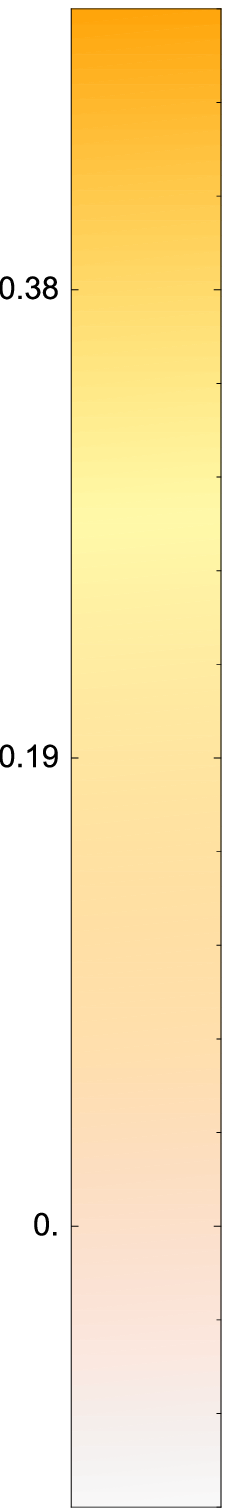}\hspace{2mm}
\includegraphics[width=0.40\linewidth]{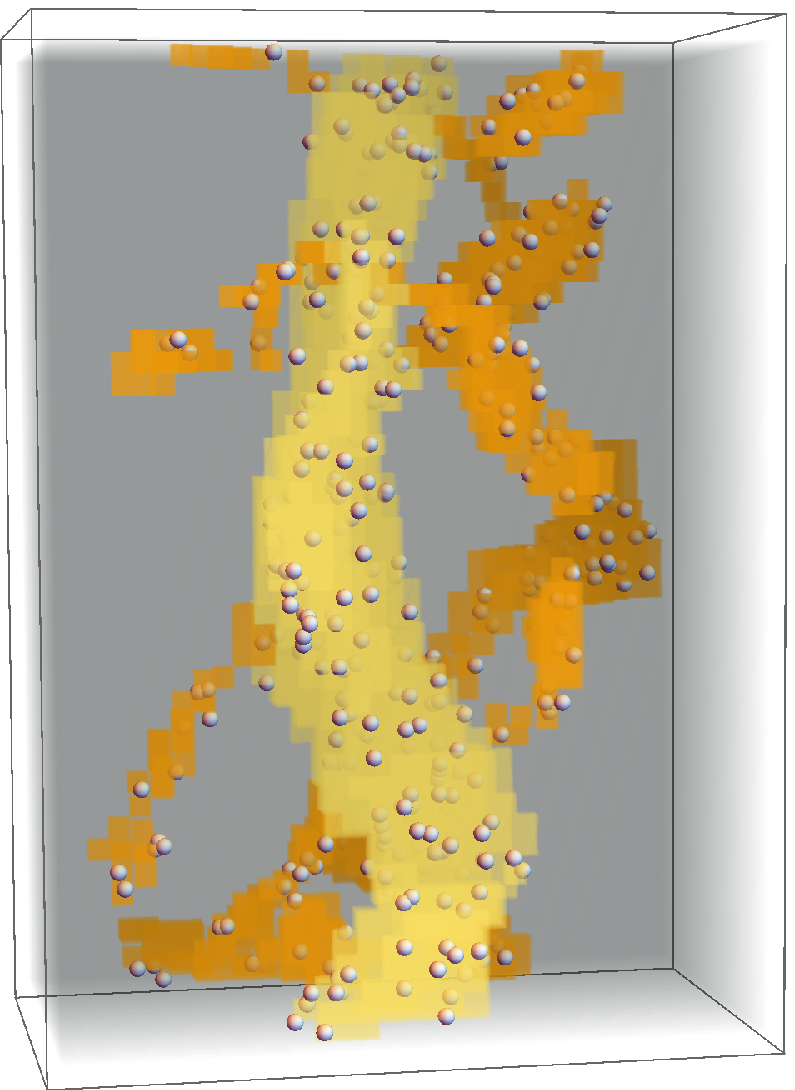}
\includegraphics[width=0.40\linewidth]{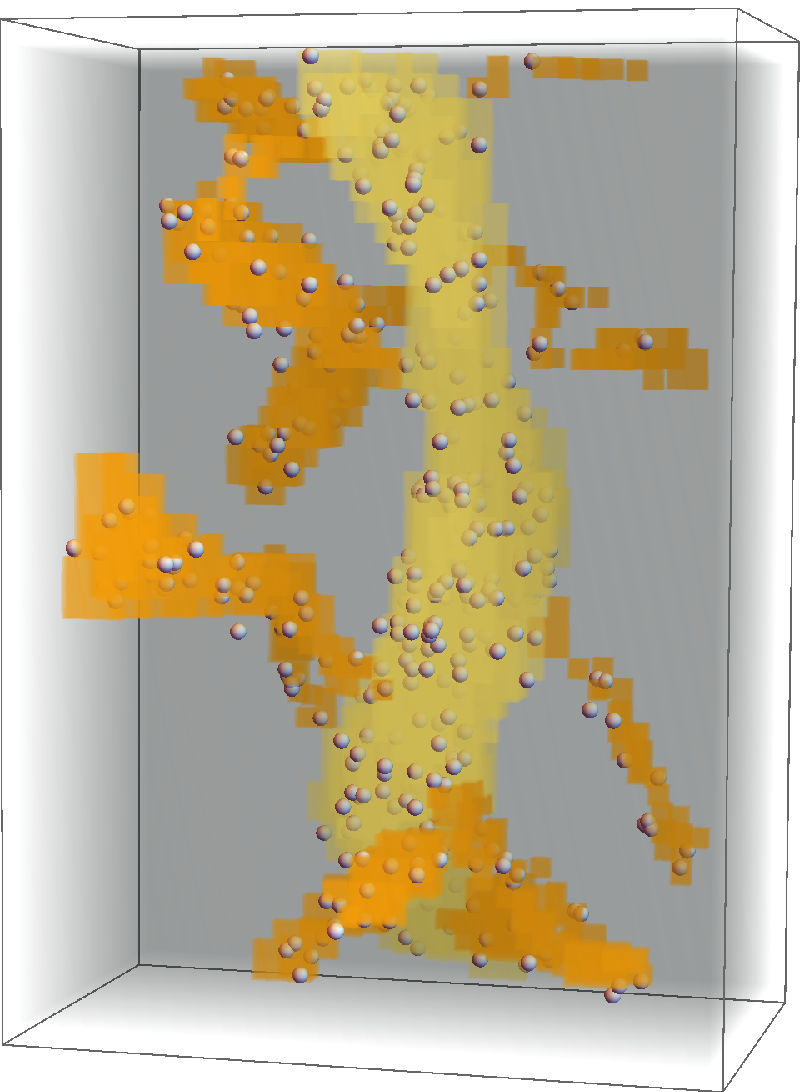}\vspace{4mm}\\
\includegraphics[height=4.5cm]{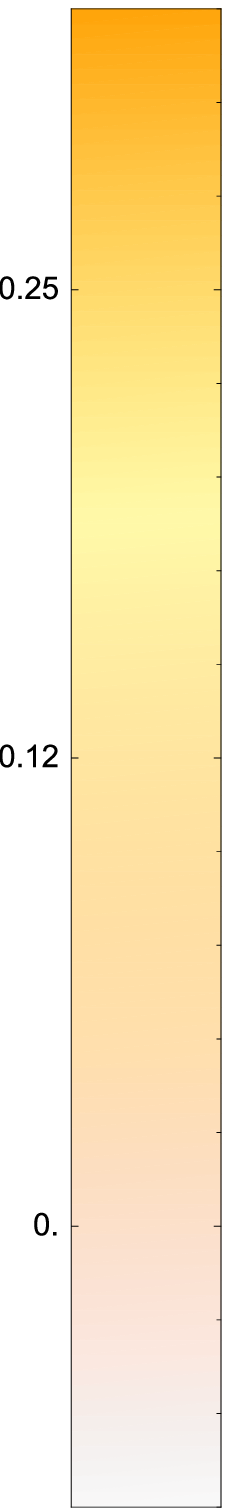}\hspace{2mm}
\includegraphics[width=0.4\linewidth]{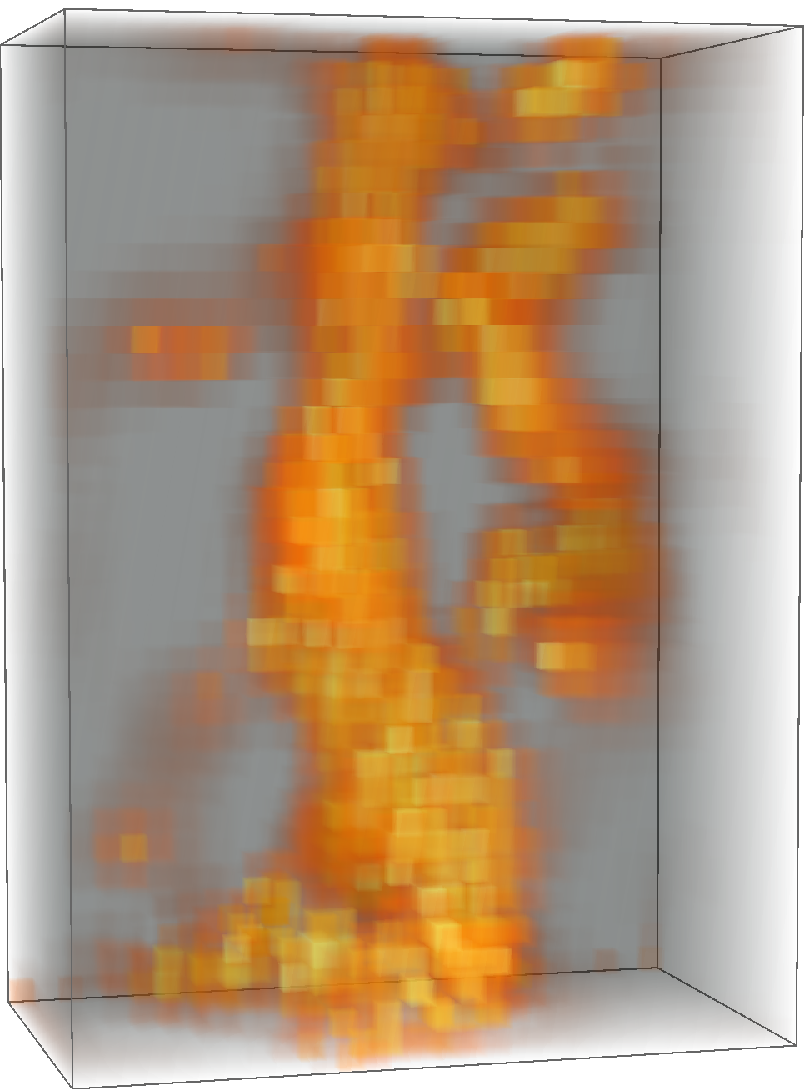}
\includegraphics[width=0.4\linewidth]{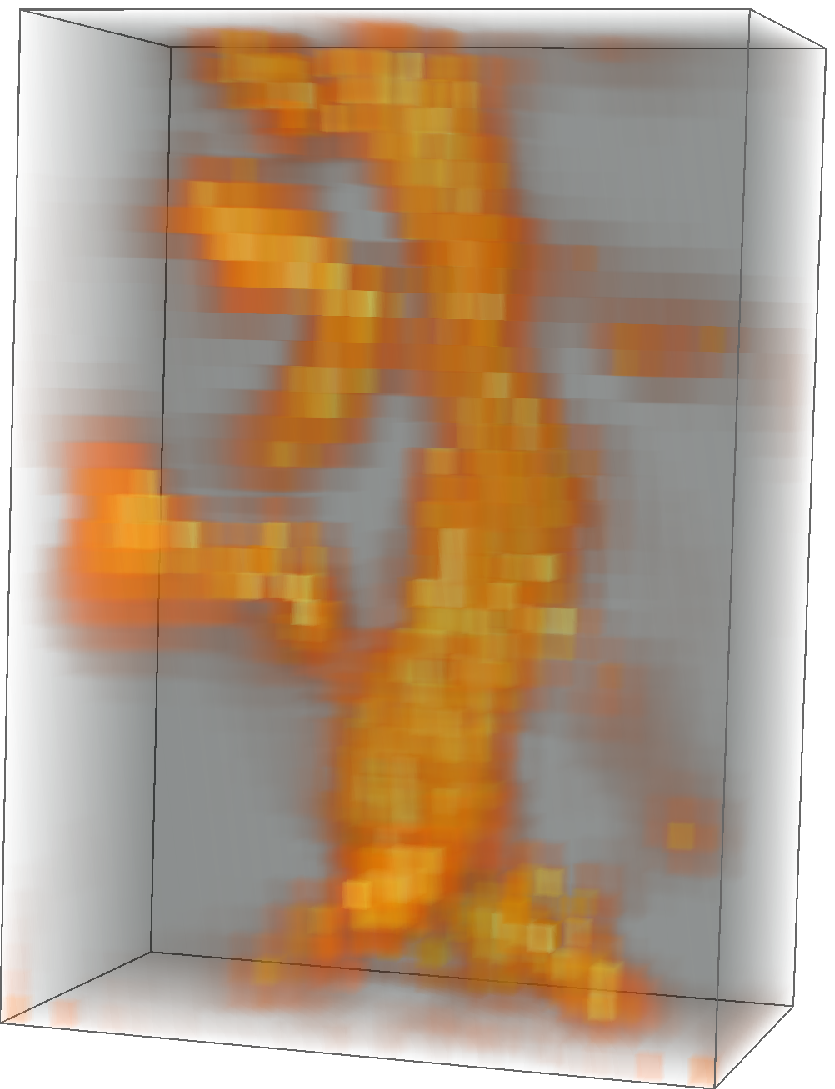}
\caption{(Color online) Two views of the model neuron (top row) and of the reconstructed neuron (bottom row). The top row also shows the positions of the sources, which are represented by gray spheres. The field of view in each image is 491 nm $\times$ 680 nm.
}
\label{fig:neurons}
\end{figure}

In Fig.~\ref{fig:neurons} we present a three-dimensional reconstruction of a spiny neuron. The image of the neuron was obtained from experiment and manually segmented \cite{ccdb,doi:10.1179/his.2000.23.3.261}. The segmented image was then voxelized onto a 491 nm $\times$ 680 nm $\times 281$ nm grid with cubic voxels of side length $15$ nm. The forward problem was solved on the above grid with biologically realistic  contrast.  The dendrite (main trunk) has contrast $\delta\varepsilon=0.25$, while the spines (branches) have contrast $\delta\varepsilon=0.38$.  Five hundred sources were placed randomly within the neuron. A grid of 15 $\times$ 15 evenly spaced detectors collected the scattered field, again spanning 120 degrees in the lateral directions.  This arrangement corresponds to $112,000$  measurements to which 1\% noise was then added. The image was reconstructed on a grid of cubic voxels with $16$ nm sides, resulting in a  overdetermined problem with 23,994 unknowns. Reconstructions from two different orientations are shown. We observe that the spines are reconstructed quite accurately. As may be expected, the dendrite is not recovered with equal fidelity. 

\begin{figure}[b]
\centering
\includegraphics[width=0.90\linewidth]{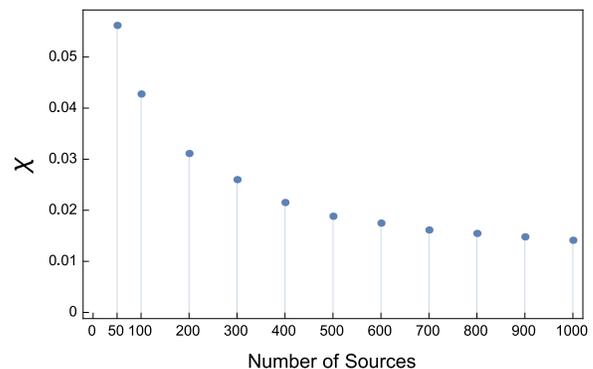}
\caption{\label{fig:phi_err}Dependence of the relative error $\chi$ on the number of sources.}
\end{figure}

In the proposed method, as in PALM itself, information from the sources is acquired serially, while the detectors are read out in parallel. It follows that in order to minimize the data collection time, it is important to determine the optimal number of sources for a given experiment. We are thus led to define the relative error 
\begin{equation}
\chi^2 = \frac{\displaystyle\sum_{\Br_1,\Br_2}\left|A(\Br_1,\Br_2)-A_{\rm rec}(\Br_1,\Br_2)\right|^2}{\displaystyle\sum_{\Br_1,\Br_2}\left|A(\Br_1,\Br_2)\right|^2} \ ,
\end{equation}
where $A$ is the measured scattering amplitude, $A_{\rm rec}$ is the scattering amplitude calculated from the reconstructed scattering potential, and the sum is carried out over all sources and detectors. In Fig.~\ref{fig:phi_err} we plot the dependence of $\chi^2$ on the number of sources used in the reconstruction of the neuron. We see that the error decreases slowly and that 500 sources is nearly optimal. 

We now examine the mathematical structure of the inverse problem. We begin by differentiating the scattering amplitude $A(\Br_1,\Br_2)$ with respect to $\Br_1$:
\begin{equation}
\label{diff}
\lap_{\Br_1}A = e^{-ik\hat\Br_2\cdot\Br_1}V(\Br_1) -k^2\int_\Omega d^3r e^{-ik\hat\Br_2\cdot\Br}G(\Br_1,\Br)V(\Br) \ ,
\end{equation}
where we have used (\ref{def_A}) and the fact that the Green's function obeys the equation $\lap_{\Br}G(\Br,\Br') + k^2G(\Br,\Br')=-\delta(\Br-\Br')$. Making use of
(\ref{def_A}) once again, we can solve (\ref{diff}) for $V$, thereby obtaining
the inversion formula
\begin{equation}
\label{inversion}
V(\Br_1) = e^{ik\hat\Br_2\cdot\Br_1}\left(\lap_{\Br_1} A(\Br_1,\Br_2) + k^2 A(\Br_1,\Br_2)\right) \ .
\end{equation}
We observe that for fixed $\Br_2$, the above result allows for a \emph{local} reconstruction of $V$. Moreover, there is in principle no limit to the resolution of the reconstruction, beyond that imposed by the accuracy of the forward model. It follows immediately that $V$ can be reconstructed with Lipschitz stability. That is, errors in $A$ propagate linearly to errors in $V$, and therefore the ISP is well-posed~\cite{natterer}. More precisely, suppose that $A$ and $A'$ are the scattering amplitudes corresponding to the potentials $V$ and $V'$ , respectively. We then have the stability estimate
\begin{equation}
\|V-V'\|_{L^2(\Omega)} \le C \|A-A'\|_{H^2(\Omega)} \ ,
\end{equation}
where $C$ is a constant that depends only on $\Omega$. Note that the presence of the Sobolev norm $\|\cdot\|_{H^2(\Omega)}$ implies the loss of some smoothness in the reconstructed potential. In contrast, we note that solving the inverse problem with scattering data obtained from external sources leads to inversion of the Laplace transform, a severely ill-posed problem~\cite{epstein}.

We close with several remarks. (i) The proposed method can be applied to refine the localization of sources in PALM by a two-step iterative procedure. In the first step, the scattering potential is determined using source locations that are determined by PALM imaging. In the second step, the positions of the sources are determined by solving the inverse source problem for the medium with the reconstructed potential. The process is then repeated until convergence. Numerical studies of this algorithm will be reported elsewhere. (ii) The inverse problem investigated here requires knowledge of the optical phase. While feasible, measurements of the phase are technically challenging. Thus it would be of some interest to develop a phaseless inversion procedure. Results in this direction have been reported for near-field inverse scattering based on measurement of the power extinguished from the incident field~\cite{carney_2001,govyadinov_2009}. The generalized optical theorem~\cite{carney_2004again} plays a crucial role in such problems and could be adapted to the inverse problem with internal sources. (iii) We have utilized the scalar theory of the optical field in this work. In future work, we plan to employ the full vector theory of electromagnetic scattering and study the inverse problem in this setting. (iv) The inversion formula (\ref{inversion}) is mainly of theoretical interest. In contrast to the algebraic inversion method that was implemented in our numerical simulations, it does not account for sampling or limited data. In principle, the scattering amplitude could be smoothed before it is numerically differentiated, leading to an alternative reconstruction method. This approach would allow for a natural way to combine the data obtained from multiple detectors, albeit with the loss of the local nature of the reconstruction. We note that the resolution would then be set by the scale over which the smoothing is carried out, rather than the optical wavelength. (v) The inverse problem has been studied within the accuracy of the FBA. The FBA holds for small weakly-scattering media, conditions that are often met in practice. Nevertheless, nonlinear corrections to the inversion formula (\ref{inversion}) are of interest and could be obtained by inversion of the Born series~\cite{inv_born}.

In conclusion, we have developed a method for reconstructing the optical properties of an inhomogeneous scattering medium with subwavelength resolution from internal sources. The required measurements may be obtained from an interferometric PALM imaging experiment. Our results were illustrated by numerical simulations that demonstrate an achievable resolution of $\lambda/25$. The associated inverse scattering problem was analyzed mathematically and shown to be well-posed. Finally, we note that concepts we have presented are quite general since they can be applied to imaging with any wave field generated by an internal source. 

\acknowledgments
This work was supported in part by the NSF grants CCF-1161233 to ACG and DMS-1619907 to JCS.



\begin{thebibliography}{10}
\bibitem{remark}
That is, the scattering potential can be obtained by differentiation of the measured field. This follows immediately from (\ref{wave_eqn}), whereby we obtain the inversion formula $\epsilon(\Br) = -\lap u(\Br)/k^2 u(\Br)$, which holds outside of the source.
\end{thebibliography}
\end{document}